\documentclass[twocolumn,pra,aps,showpacs,superscriptaddress]{revtex4-1}
\usepackage[utf8]{inputenc}
\usepackage{url,psfrag,graphicx}
\usepackage{dcolumn}
\usepackage{amsmath,amssymb}
\usepackage{bm}
\usepackage{pstricks}
\usepackage{hyperref}
\usepackage{epsfig,placeins}

\def\<{\left<}
\def\>{\right>}
\def\ket|#1>{\left|#1\right>}
\def\bra<#1|{\left<#1\right|}
\def\elem<#1|#2|#3>{\left<#1\right|#2\left|#3\right>}
\def\({\left(}
\def\){\right)}

\begin{document}

\title[Short Title]{Entanglement in correlated random spin chains, RNA
  folding and kinetic roughening}

\author{Javier Rodr\'{\i}guez-Laguna}
\affiliation{Departamento de F\'{\i}sica Fundamental, UNED, Madrid, Spain.}

\author{Silvia N. Santalla} 
\affiliation{Departamento de F\'{\i}sica, Universidad Carlos III de
  Madrid, Legan\'es, Spain.}

\author{Giovanni Ram\'{\i}rez}
\affiliation{Instituto de F\'{\i}sica Te\'orica UAM/CSIC, Madrid, Spain.}

\author{Germ\'an Sierra}
\affiliation{Instituto de F\'{\i}sica Te\'orica UAM/CSIC, Madrid, Spain.}

\date{January 13, 2016}

\begin{abstract}
Average block entanglement in the 1D XX-model with uncorrelated random
couplings is known to grow as the logarithm of the block size, in
similarity to conformal systems. In this work we study random spin
chains whose couplings present long range correlations, generated as
gaussian fields with a power-law spectral function. Ground states are
always planar valence bond states, and their statistical ensembles are
characterized in terms of their block entropy and their bond-length
distribution, which follow power-laws. We conjecture the existence of
a critical value for the spectral exponent, below which the system
behavior is identical to the case of uncorrelated couplings. Above
that critical value, the entanglement entropy violates the area law
and grows as a power law of the block size, with an exponent which
increases from zero to one. Similar planar bond structures are also
found in statistical models of RNA folding and kinetic roughening, and
we trace an analogy between them and quantum valence bond
states. Using an inverse renormalization procedure we determine the
optimal spin-chain couplings which give rise to a given planar bond
structure, and study the statistical properties of the couplings whose
bond structures mimic those found in RNA folding.
\end{abstract}


\maketitle


\section{Introduction}

Entanglement in disordered spin chains has received much attention
recently \cite{Refael.04, Laflorencie.05, Hoyos.07, Ramirez.14}. The
main reason is that, as opposed to on-site disordered systems
\cite{Anderson.58}, long-distance correlations are not destroyed in
this case, but only modified in subtle ways. Thus, for the 1D
Heisenberg and XX models with uncorrelated random couplings, the von
Neumann entropy of blocks of size $\ell$ is known to violate the area
law and grow as $\log(\ell)$, similarly to the conformal case
\cite{Vidal.03,Eisert.10}. The prefactor, nonetheless, is different:
it is still proportional to the the central charge of the associated
conformal field theory (CFT), but multiplied by an extra $\log(2)$
factor. Moreover, the R\'enyi entropies do not satisfy the predictions
of CFT \cite{Cardy.09}, because these models are not conformal
invariant.

A very relevant tool of analysis is the strong disorder
renormalization group (SDRG) devised by Dasgupta and Ma
\cite{Dasgupta.80}, which shows that the ground state of Heisenberg or
XX chains with strong disorder can be written as a product of {\em
  random singlets}, in which all spins are paired up making SU(2)
singlet bonds. Furthermore, the renormalization procedure prevents the
bonds from crossing, i.e., the bond structure will always be {\em
  planar}. The paired spins are often neighbours, but not always. As
it was shown \cite{Fisher.95,Refael.04}, the probability distribution
for the singlet bond lengths, $P_B(\ell)$ falls as a power-law,
$P_B(\ell)\sim \ell^{-\eta}$, with $\eta=2$. Entanglement of a block
can be obtained just by counting the number of singlets which must be
cut in order to isolate the block, and multiplying by the entanglement
entropy of one bond, which is $\log(2)$.

Under the SDRG flow, the variance of the couplings increases and its
correlation length decreases, thus approaching the so-called {\em
  infinite randomness fixed point} (IRFP) \cite{Fisher.95}. Is this
fixed point unique? Not necessarily. If the couplings present a
diverging correlation length, we might have other fixed points of the
SDRG. For example, if the couplings decay exponentially from the
center, they give rise to the {\em rainbow phase}, in which singlets
extend concentrically \cite{Vitagliano.10, Ramirez.14.2,
  Ramirez.15}. In that case, all couplings are correlated. But we may
also devise ensembles of couplings which present long-range
correlations, but are still random.

A glimpse of some further fixed points can be found by observing the
statistical mechanics of the secondary structure of RNA
\cite{Tinoco.99}. A simple yet relevant model is constituted by a
closed 1D chain with an even number of RNA bases, which we call sites,
which are randomly coupled in pairs with indices of different parity
\cite{Wiese.06,Wiese.08}. Each pair constitutes an RNA bond, and the
only constraint is that no bonds can cross. Therefore, the ensemble of
secondary structures of RNA can be described in terms of planar bond
structures, just like ground states of disordered spin-chains. Wiese
and coworkers \cite{Wiese.08} studied the probability distribution for
the bond lengths, and found $P_B(l)\sim l^{-\eta}$, with
$\eta=(7-\sqrt{17})/2 \approx 1.44$.

Furthermore, the studies of RNA folding included a very interesting
second observable. The planar bond structure can be mapped to the
height function of a discretized interface \cite{Wiese.08}. We can
define the expected roughness of windows of size $\ell$, $W(\ell)$, as
the deviation of the height function over blocks of size $\ell$, which
can be shown to scale in RNA folding structures like $W(\ell)\approx
\ell^\alpha$, with $\alpha=(\sqrt{17}-3)/2\approx
0.56$. Interestingly, $\eta+\alpha=2$.

As we will show, the interface roughness is very similar to the
entanglement entropy of blocks of size $\ell$, and they are
characterized by similar exponents. In the IRFP phase for random
singlets, notice that the entropy is characterized by a zero exponent,
due to the logarithmic growth, and $\eta=2$. Therefore, it is also
true that $\eta+\alpha=2$. We may then ask, what is the validity of
this scaling relation? Does the RNA folding case correspond to some
choice of the ensemble of coupling constants for a spin-chain? Can we
obtain other fixed points which interpolate between the IRFP and the
RNA folding cases?

We may keep in mind that the couplings in some spin chain models
(e.g., the XX model) can be mapped into modulations of the space
metric \cite{Boada.11}. Thus, we are obtaining, in a certain regime,
the relation between the statistical structure of the space metric and
the statistical properties of entanglement of the vacuum, i.e., the
ground state of the theory. 

This article is organized as follows. Section \ref{sec:model}
introduces our model and the renormalization procedure used throughout
the text. Moreover, it discusses the consequences of the planarity of
the pairing structures which characterize the states. In section
\ref{sec:results} we establish our strategy to sample highly
correlated values of the couplings, and show numerically the behavior
of the entropy and other observables. In section \ref{sec:rna} we
focus on the relation between the RNA folding problem and our
disordered spin chains, and determine an inverse algorithm to compute
a parent Hamiltonian for any planar state, exemplifying it with the
RNA folding states. How generic are planar states is the question
addressed in section \ref{sec:generic}, showing that they are
non-generic through the study of their entanglement entropy. The
article ends in section \ref{sec:conclusions} discussing our
conclusions and ideas for further work.


\section{Disordered Spin Chains and Planar States}
\label{sec:model}

Let us consider for simplicity a spin-1/2 XX chain with $N$ (even)
sites and periodic boundary conditions, whose Hamiltonian is

\begin{equation}
H=-\sum_{i=1}^N J_i \( S^x_i S^x_{i+1} + S^y_i S^y_{i+1} \)
\label{eq:ham}
\end{equation}
where the $J_i$ are the coupling constants, which we will assume to be
positive and strongly inhomogeneous. More precisely, we assume that
neighboring couplings are very different. Notice that we do not impose
them to be random.

In order to obtain the ground state (GS), we can employ the {\em
  strong disorder renormalization group} (SDRG) method of Dasgupta and
Ma \cite{Dasgupta.80}. At each renormalization step, we pick the
maximal coupling, $J_i$, decimate the two associated spins, $i$ and
$i+1$, and establish a {\em singlet bond} between them. The
neighboring sites are then joined by an effective coupling given by
second order perturbation theory:

\begin{equation}
\tilde J_i = J_{i-1}J_{i+1}/J_i
\label{eq:dasguptama}
\end{equation}
among the next neighbours of the link, $i-1$ and $i+2$. It is
convenient to use a set of auxiliary variables, that we will call {\em
  log-couplings}: $t_i=-\log(J_i)$. The main reason is that, for them,
the Dasgupta-Ma renormalization rule becomes {\em additive}:

\begin{equation}
\tilde  t_i=t_{i-1}+t_{i+1}-t_i
\label{eq:linearrg}
\end{equation}

Once the SDRG procedure is finished we can read our GS as a product
state of {\em singlet} valence bonds. 

\begin{equation}
\ket|\Psi_{GS}>= \prod_{(i,j)\in {\cal P}} {1\over\sqrt{2}} \(
\ket|+->_{ij} - \ket|-+>_{ij} \) 
\label{eq:state}
\end{equation}
where ${\cal P}$ denotes a set of $N/2$ pairing bonds among the $N$
spins. Many properties of these ground states have been studied in the
last thirty years \cite{Fisher.95, Refael.04, Laflorencie.05,
  Hoyos.07, Vitagliano.10, Fagotti.11, Ramirez.14, Ramirez.14.2,
  Ramirez.15}. One of the most salient of those properties is the fact
that the pairing ${\cal P}$ which results from the renormalization
procedure {\em must be planar}, i.e., it can be drawn without any two
bonds crossing. States of the form \eqref{eq:state} which fulfill this
requirement will be called from now on {\em planar states}.


\subsection{Planar Pairings}
\label{subsec:planar}

\begin{figure*}
\epsfig{file=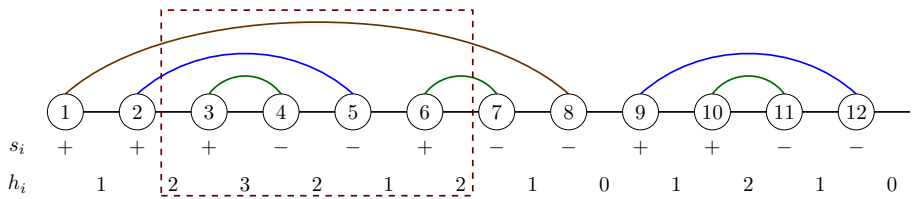,width=12cm}
\caption{Illustrating the planar pairings. A set of $N=12$ sites,
  coupled by a planar pairing. Each site gets a {\em spin} value,
  $s_i=+1$ or $-1$. The height function, $h_i$ is shown in the bottom
  row. A block is marked, containing sites 3 to 6.}
\label{fig:illust}
\end{figure*}

In more formal terms, let (even) $N$ be the number of nodes, and a
bond is defined as an ordered pair ${\bf p}=(p_1,p_2)$, where
$p_1,p_2\in {\mathbb Z}_N$ are the nodes joined. In principle,
$(p_1,p_2)\neq (p_2,p_1)$. We define the covered nodes by the bond as
$C({\bf p})\equiv \{p_1+1,\cdots,p_2-1\}$. Notice that, if $p_1$ and
$p_2$ are consecutive, $C({\bf p})=\emptyset$. Given two bonds, ${\bf
  p}$ and ${\bf q}$, we say that ${\bf p}\subset {\bf q}$ if $C({\bf
  p})\subset C({\bf q})$. If neither $C({\bf p})\subset C({\bf q})$ or
$C({\bf q})\subset C({\bf p})$, we say that the bonds {\em cross}. A
planar bond structure is defined as a set of $N/2$ bonds which {\em do
  not cross}. Thus, the bonds form a {\em nested graph}. An important
remark is that the two nodes joined by bond must have different
parity. See Fig. \ref{fig:illust} for an illustration.

Let us assume that the nodes are indexed counterclockwise. We can now
define for each node $i$ a value $s_i$ to be either $+1$ or $-1$
depending on whether it is the source or the sink of a bond, as shown
in Fig. \ref{fig:illust}. Of course, the sum of all the $s_i$ around
the full system should be zero: $\sum_{k=1}^N s_k=0$. The $s_i$ can be
considered as {\em slopes} of a height function,

\begin{equation}
h_i\equiv \sum_{k=1}^i s_k.
\label{eq:height_function}
\end{equation}

Of course, this definition is not translation invariant, since we
start counting at node 1. In order to avoid that, let $h_0$ denote the
absolute minimum of this height function. Then, we can define the {\em
  absolute height function}, $H_i\equiv h_i-h_0$. Its meaning is the
following: it denotes the number of bonds passing above the link in
the circle joining nodes $i$ and $i+1$. By construction, this absolute
height function has, at least, one zero.

\subsection{Dyck Language and Catalan Numbers}

There is a close analogy between planar pairings and the {\em Dyck
  language} \cite{Stanley.99}. A Dyck word is a string of symbols from
the alphabet $\{+,-\}$ such that the number of $+$ counted from the
left is always greater or equal to the number of $-$. Equivalently,
they are the set of {\em properly balanced} parenthesis. This means
that their height function $h_i$, as defined in
Eq. \eqref{eq:height_function}, is positive for all $i$. The
difference between our planar pairings and the Dyck language resides
entirely in the periodic boundary conditions. If, in our {\em
  circular} planar structures, we break at the absolute minimum of the
height function, the analogy with Dyck words becomes complete.

How many different planar states are there for a system of fixed size
$N=2M$? Let us denote this value by $P_M$. Disregarding the ordering
of the sites in each bond, which merely contributes a general $2^M$
factor, we can provide a recursive relation. Site 1 must be linked to
an even site, $2k$. Then it creates two {\em regions}, one of size
$2k-2$ and the other $2M-2k$. Thus, we get

\begin{equation}
P_M = \sum_{k=1}^M P_{k-1} P_{M-k}
\label{eq:recurrence}
\end{equation}
along with $P_2=1$, which is known in the literature as {\em Segner's
  recurrence} \cite{Stanley.99}, which gives rise to the Catalan
numbers:

\begin{equation}
P_M = {1\over M+1} { 2M \choose  M }
\label{eq:catalan}
\end{equation}

\subsection{Entanglement of Planar States}

Given a planar state of the form \eqref{eq:state}, we can easily
compute the entanglement entropy of any block $B$: using 2 as the base
for the logarithms, it coincides with the number of bonds which must
be cut in order to separate it from the rest of the system
\cite{Refael.04, Laflorencie.05, Ramirez.14}.

\begin{equation}
S(B)\equiv \sum_p [p_1\in B \;\oplus\; p_2\in B] 
\label{eq:define_entropy}
\end{equation}
where $\oplus$ stands for the {\em exclusive or} (xor) symbol, which
means that {\em either} $p_1\in B$ or $p_2\in B$, but not both. We can
prove the following theorem which relates the height function and
entanglement. Let $[i..j]$ denote the block $\{i,\cdots,j\}$. Then,

\begin{equation}
S(B_{[i..j]}) = H_{i-1} + H_j - 2\min_{k\in \{i-1 \cdots j\}} H_k.
\label{eq:entropy_height}
\end{equation}

The meaning of that equation is the following. $H_{i-1}$ represents
the bonds that {\em enter} the block from its left end, and $H_j$ the
bonds which exit from its right. For an example, see
Fig. \ref{fig:illust}. The block marked with the dashed box is
$B_{[3..6]}$. The number of bonds entering from the left is $H_2=2$,
and the number of bonds leaving from the right is $H_6=2$.  But not
all those bonds contribute to the entropy. Some of them just {\em fly
  over} the block, and we can separate the block without touching
them. Let $h_F$ be the number of those flying bonds, in our example
$h_F=1$, the bond from site 1 to site 8. The links entering from the
left, $H_{i-1}$ are either overflying ($h_F$) or not ($h_L$):
$H_{i-1}=h_F+h_L$. Similarly, on the right we have $H_j=h_F+h_R$, and
the block entropy is given by $h_L+h_R$. We will proceed to prove that
$h_F$ is given by the minimum of the height function inside the
block. Since the bonds which contribute to the entropy, $h_L$ and
$h_R$, do not {\em fly} over the block, they must either {\em end}
inside it ($h_L$) or {\em start} inside it $h_R$. Since the bonds can
not cut, the $h_L$ bonds from the left must have ended {\em before}
any of the $h_R$ start. At that very moment, only the {\em flying}
bonds will remain. In Fig. \ref{fig:illust}, this moment takes place
between sites 5 and 6, $h_L=1$, $h_R=1$ and the block entropy is
$S=2$. Thus, the minimum value of the height function is, exactly,
$h_F$. We have $H_{i-1}+H_j-2h_F = h_L+h_R =S$, as required.

Notice that we can rewrite expression \eqref{eq:entropy_height} as
$S(B_{[i..j]})=(H_{i-1}-H_{min}) + (H_j - H_{min})$, thus showing a
connection between the block entropy and the {\em average variation}
of the height within the block, i.e. the roughness of the
interface. The main difference is that the entanglement entropy gives
special relevance to the boundaries.


\section{Correlated random spin chains}
\label{sec:results}

The statistical properties of the ground states of Hamiltonians of the
form \eqref{eq:ham} when the couplings $\{J_i\}$ are picked randomly
and {\em uncorrelated} have been determined in a series of papers
\cite{Dasgupta.80, Fisher.95, Refael.04, Laflorencie.05, Hoyos.07,
  Fagotti.11, Ramirez.14}. The SDRG procedure converges to the
so-called {\em infinite randomness fixed point} (IRFP). Along the RG,
the variance of the effective couplings grow, and their correlation
length decreases. It has been shown that the average entanglement
entropy of a block of size $\ell$ follows the expression
\cite{Fagotti.11, Ramirez.14}:

\begin{equation}
S(\ell) \approx \log(2)\; {c\over 3}\; \log\( {N\over \pi} Y\(
{\pi\ell\over N} \) \) + c',
\label{eq:s_rxx}
\end{equation}
where $Y(x)$ is a scaling function and $c$ is the central charge of
the associated CFT, i.e., the one which corresponds to the homogeneous
(conformal) case, with all the $J_i$ equal. In our case,
$c=1$. Surprisingly, expression \eqref{eq:s_rxx} is very similar to
the conformal expression \cite{Vidal.03}:

\begin{equation}
S_{CFT}(\ell) \approx {c\over 3}\; \log\( {N\over \pi} \sin\(
{\pi\ell\over N} \) \) + c'.
\label{eq:s_cft}
\end{equation}
The scaling function $Y(x)$ is, in fact, rather similar to $\sin(x)$
\cite{Fagotti.11,Ramirez.14}. 

Another relevant observable which helps characterize the IRFP is the
{\em bond-length} probability, i.e., given a singlet bond $(i,j)$,
determine the probability distribution for its length $l=|i-j|$,
$P_B(l)$. This value is directly related to the two-point correlation
function \cite{Fisher.95}. In the uncorrelated case, it is known to
behave, for $l\ll N$, as a power-law: $P_B(l)\approx l^{-\eta}$, where
$\eta=2$ \cite{Fisher.95}.

\subsection{Correlated couplings}

Our aim is to characterize the ground states of Hamiltonian
\eqref{eq:ham} when the couplings $J_i$ are random, but present
non-trivial correlations. If these correlations are short ranged, they
will be washed away by the renormalization procedure, and return to
the IRFP. Thus, we will consider the case of long-range correlations.

Let us establish a procedure to obtain samples from sets of
log-couplings $\{t_i\}$ which present long-range correlations, by
employing a suitable Fourier expansion:

\begin{equation}
t_j = \sum_k A_k \sin(jk+\phi_k)
\label{eq:fourier_random}
\end{equation}
where $k$ are a set of allowed momenta, $k_n=2\pi n/N$, with $n\in
\{1,\cdots,2N\}$. We do not include moment zero, since it would amount
to a global constant which would be irrelevant for the SDRG. The
values $A_k$ and $\phi_k$ are chosen as independent random
variables. The phase $\phi_k$ is taken to be uniformly distributed in
$[0,2\pi)$ and

\begin{equation}
A_k=k^{-\gamma} u_k,
\label{eq:ak}
\end{equation}
where the $u_k$ are independent gaussian variates with zero average
and variance one, and $\gamma$ is a fixed spectral exponent.

\begin{figure}
\epsfig{file=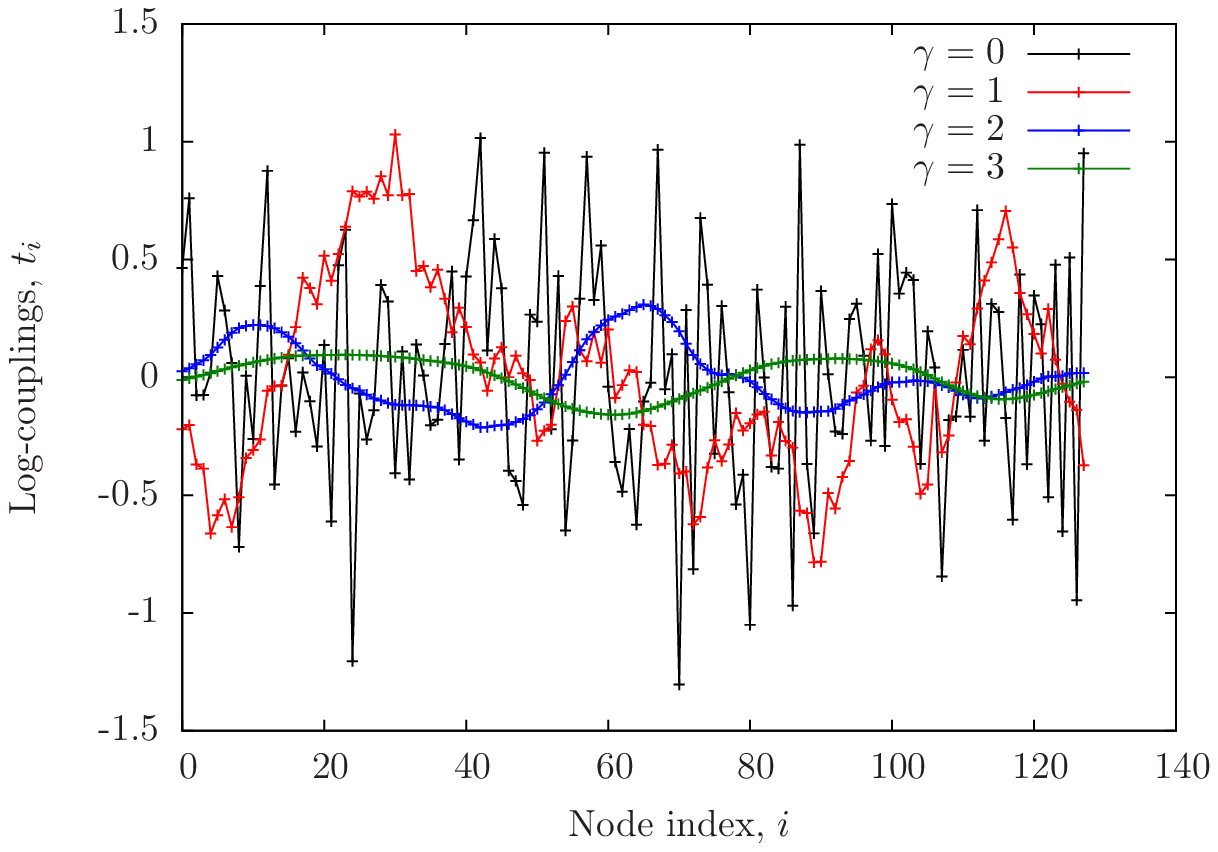,width=8cm}
\epsfig{file=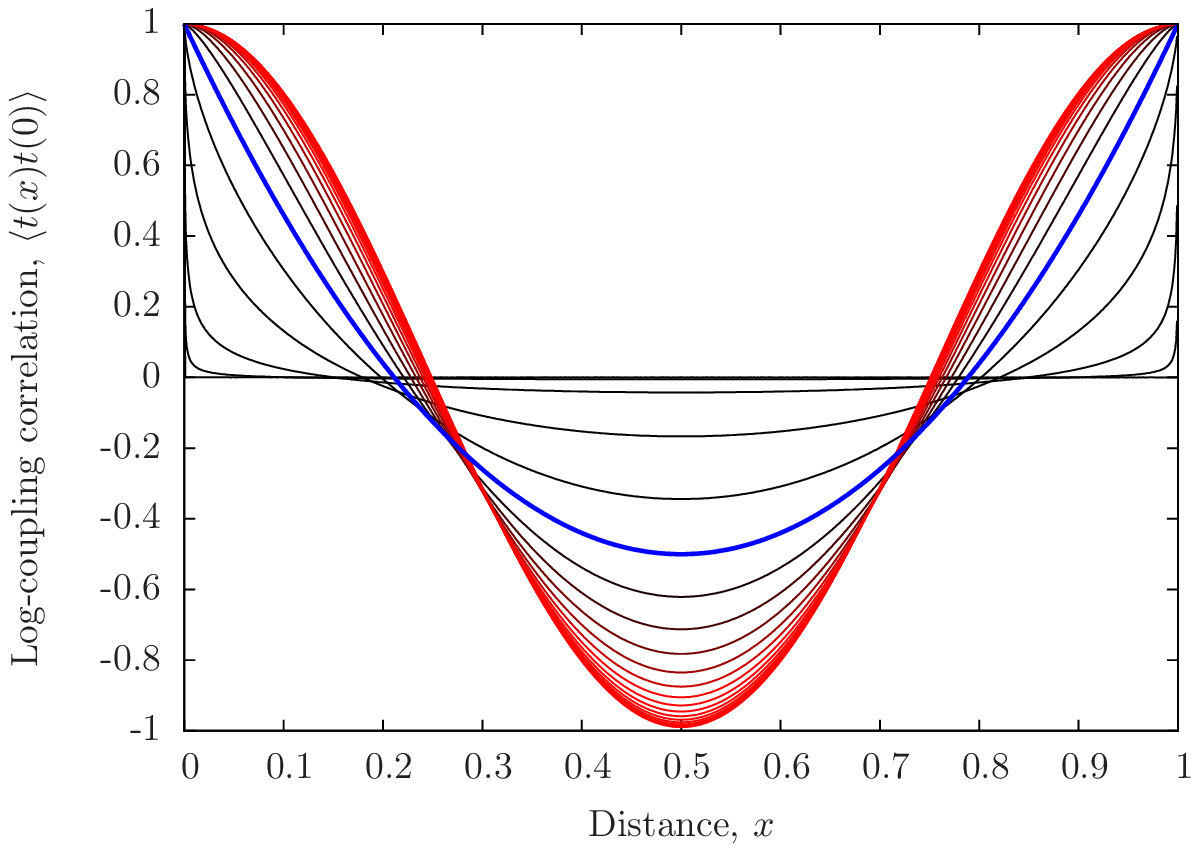,width=8cm}
\caption{(A) samples of $\{t_i\}$ for $N=128$ and increasing values of
  $\gamma$. (B) real-space correlation function for several values of
  $\gamma$ from zero (black) to infinity (red), obtained analytically
  for $N=1000$, from expression \eqref{eq:correlator_sum}. The value
  $\gamma=1$ appears remarked in blue.}
\label{fig:corrj}
\end{figure}

If $\gamma=0$, all momenta in expression \eqref{eq:fourier_random} get
the same weight, and we obtain again an {\em uncorrelated} set of
$t_i$. As we increase $\gamma$, the larger momenta get less and less
weight, and we are left with only the lowest momenta. This implies
that the set of $t_i$ have stronger correlations.
Fig. \ref{fig:corrj} (A) shows typical samples for increasing values
of $\gamma$.

The ensemble of log-couplings presents zero correlations in momentum
space, but strong correlations in real space for increasing
$\gamma$. The correlation function is translation invariant by
construction, and given by

\begin{equation}
\<t(x)t(0)\> \propto  \sum_{k\geq 2\pi/N} {1\over k^{2\gamma}} \cos(kx)
\label{eq:correlator_sum}
\end{equation}
For $N=1000$, Fig. \ref{fig:corrj} (B) shows the correlation as a
function of the distance, normalized to have a maximal value of
one. For $\gamma=0$, the correlation is identically zero for all
$x>0$. For $\gamma\to\infty$ it approaches a cosine function. The
value $\gamma=1$, which will have special relevance in the rest of the
text, appears marked. We should remark that, although
Eq. \eqref{eq:correlator_sum} makes perfect sense for all finite
values of $N$, the expression diverges in the thermodynamic limit for
$\gamma\leq 1/2$.

\begin{figure}
\epsfig{file=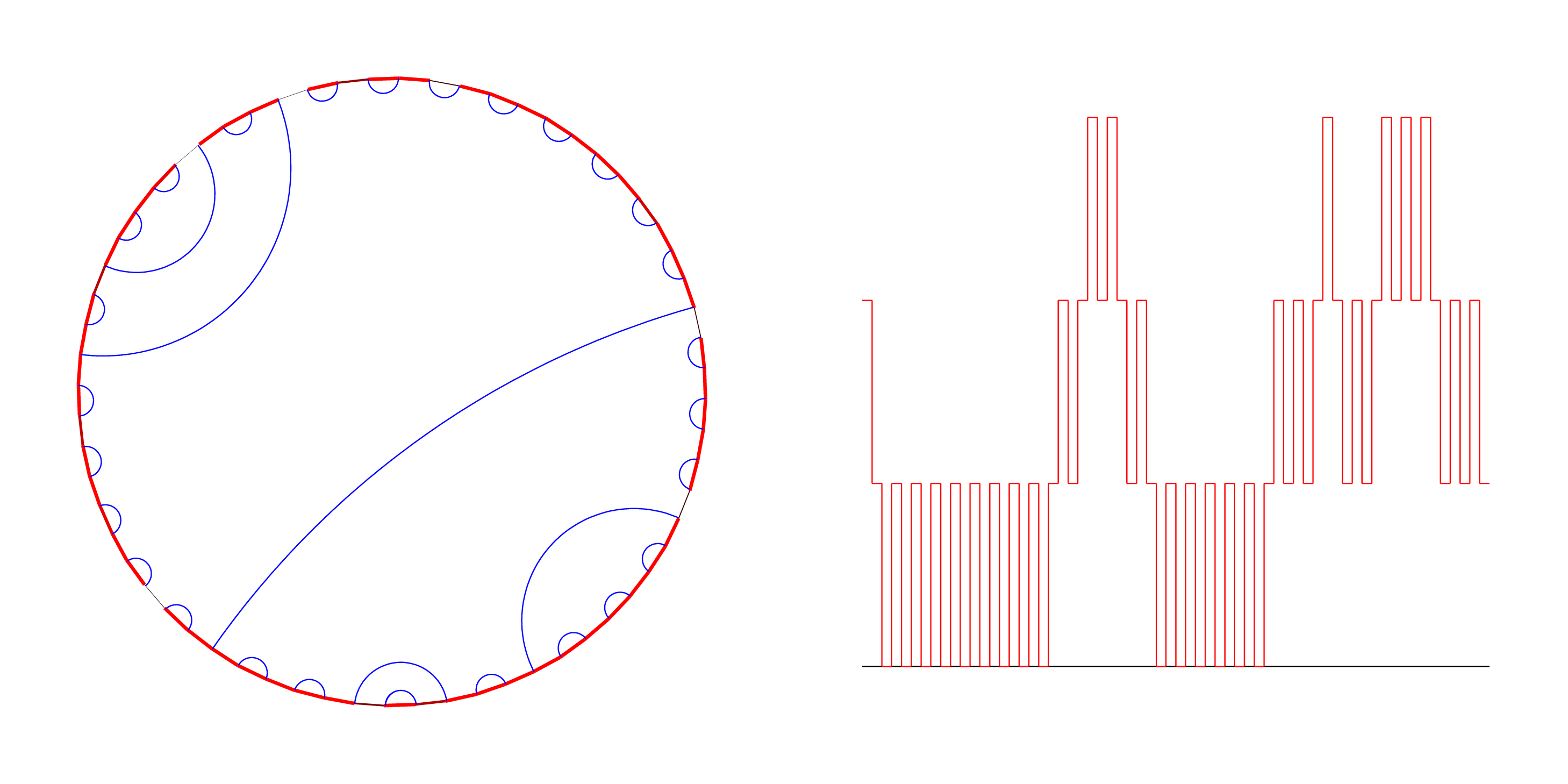,width=9cm}
\epsfig{file=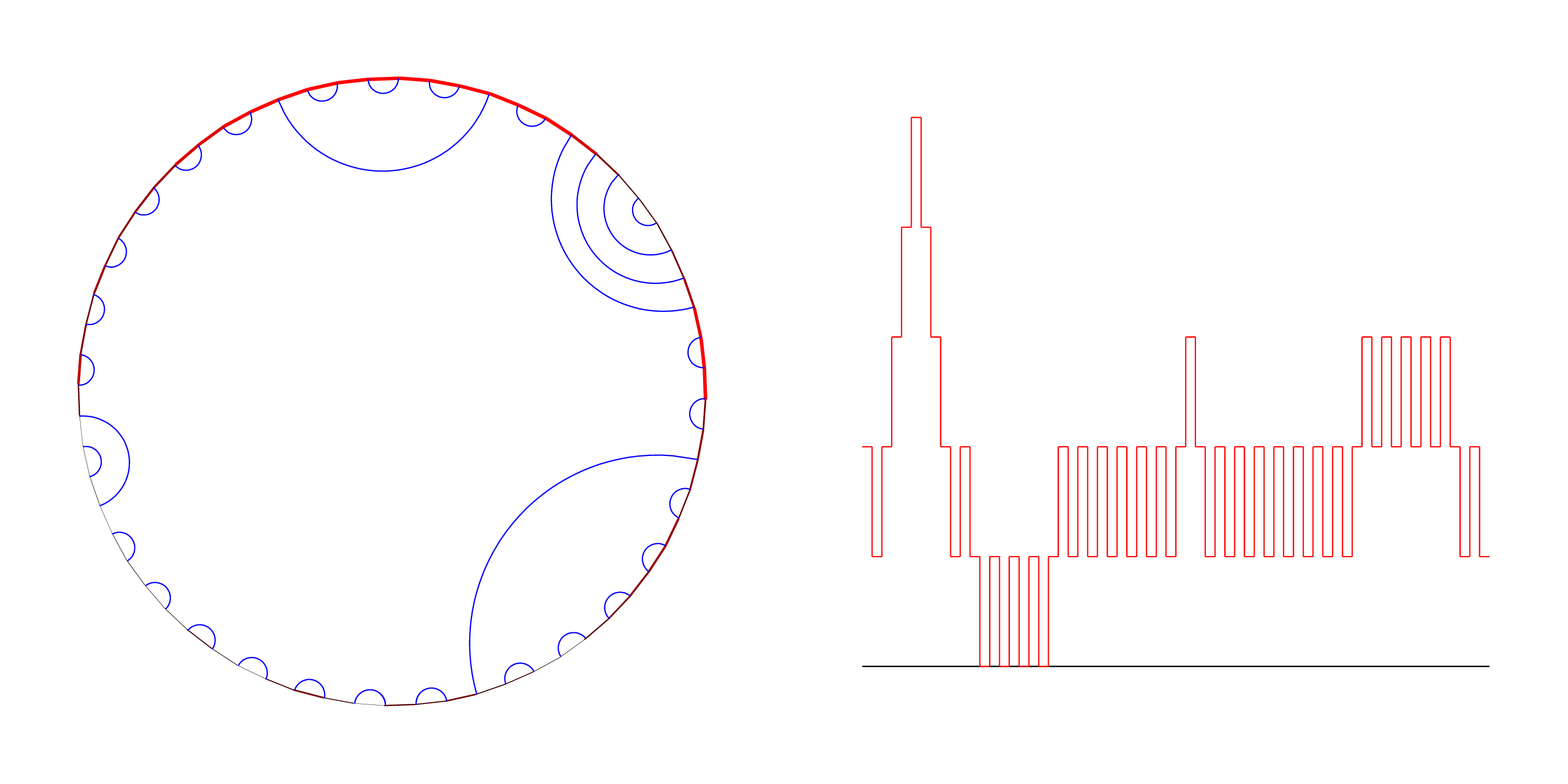,width=9cm}
\epsfig{file=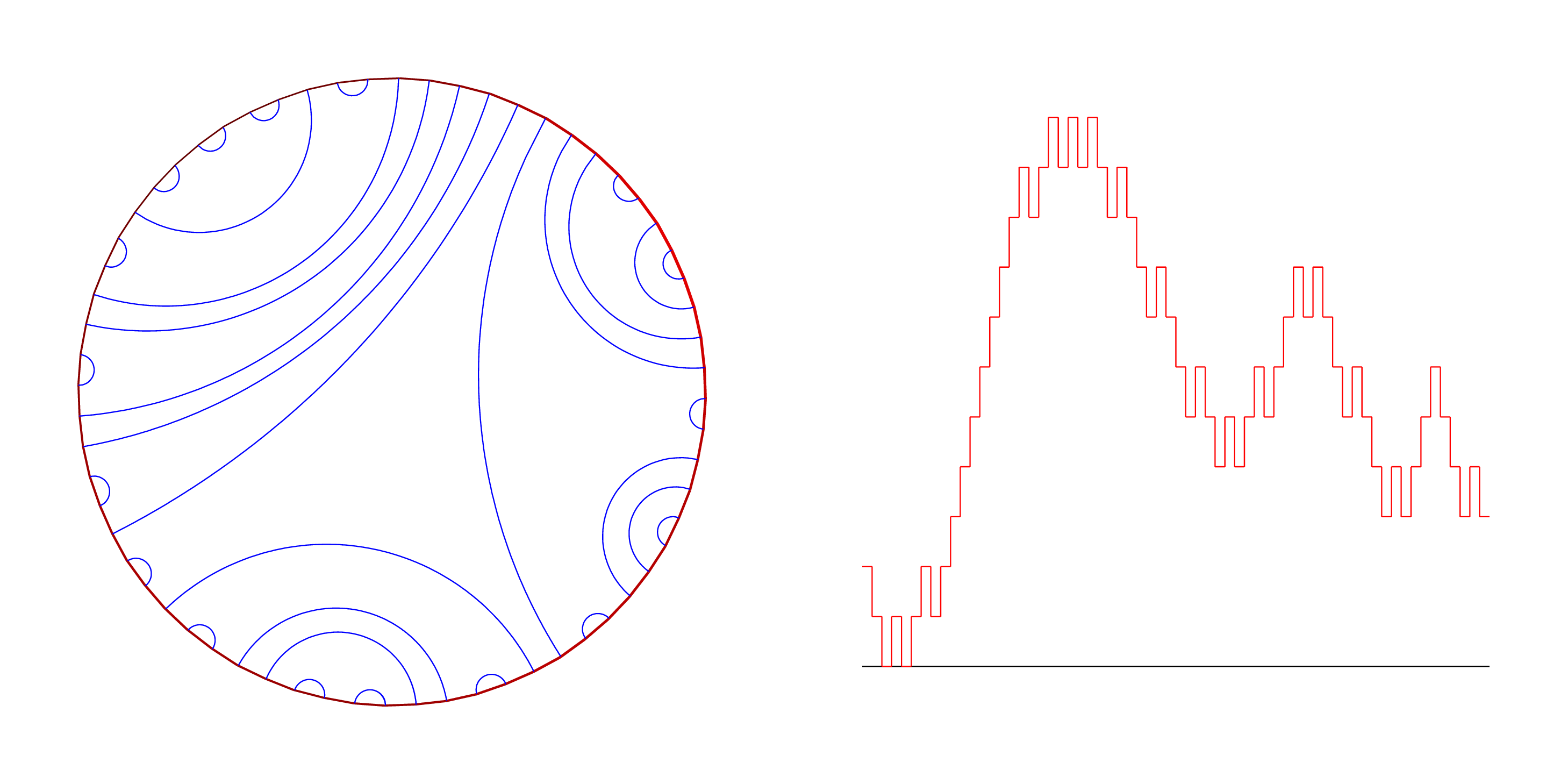,width=9cm}
\epsfig{file=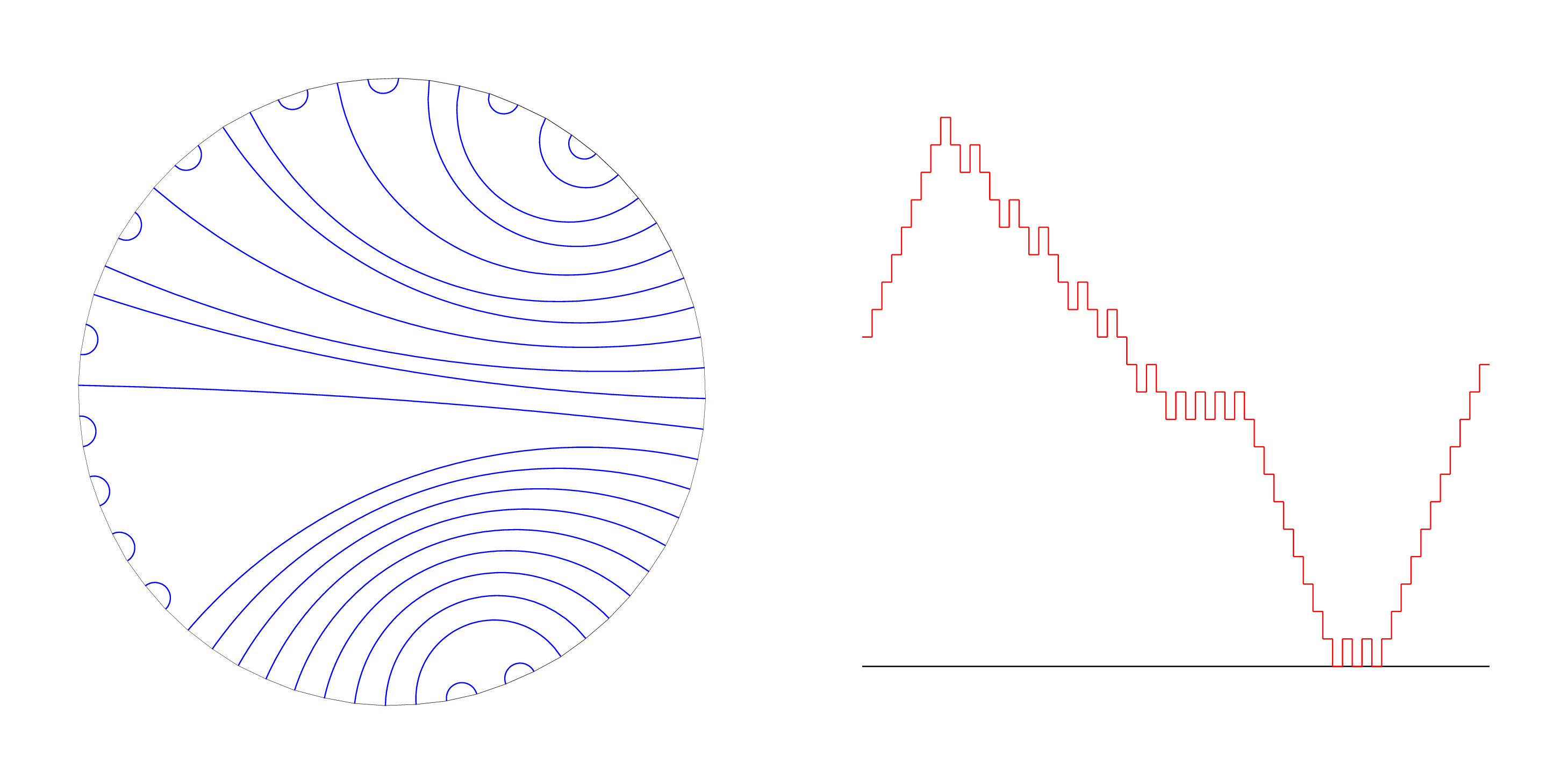,width=9cm}
\caption{Bond diagrams for different samples with $N=64$ and $\gamma$
  running (downwards) from $\gamma=0$ to $\gamma=3$. To their right,
  the corresponding height profiles.}
\label{fig:diagrams}
\end{figure}

Fig. \ref{fig:diagrams} shows some sample planar pairings for
different values of $\gamma$, in the range from $\gamma=0$ (no
correlations) to $\gamma=3$ (large correlations), along with their
corresponding height diagrams.

\subsection{Entanglement, Roughness and Bond-Lengths}

\begin{figure}
\epsfig{file=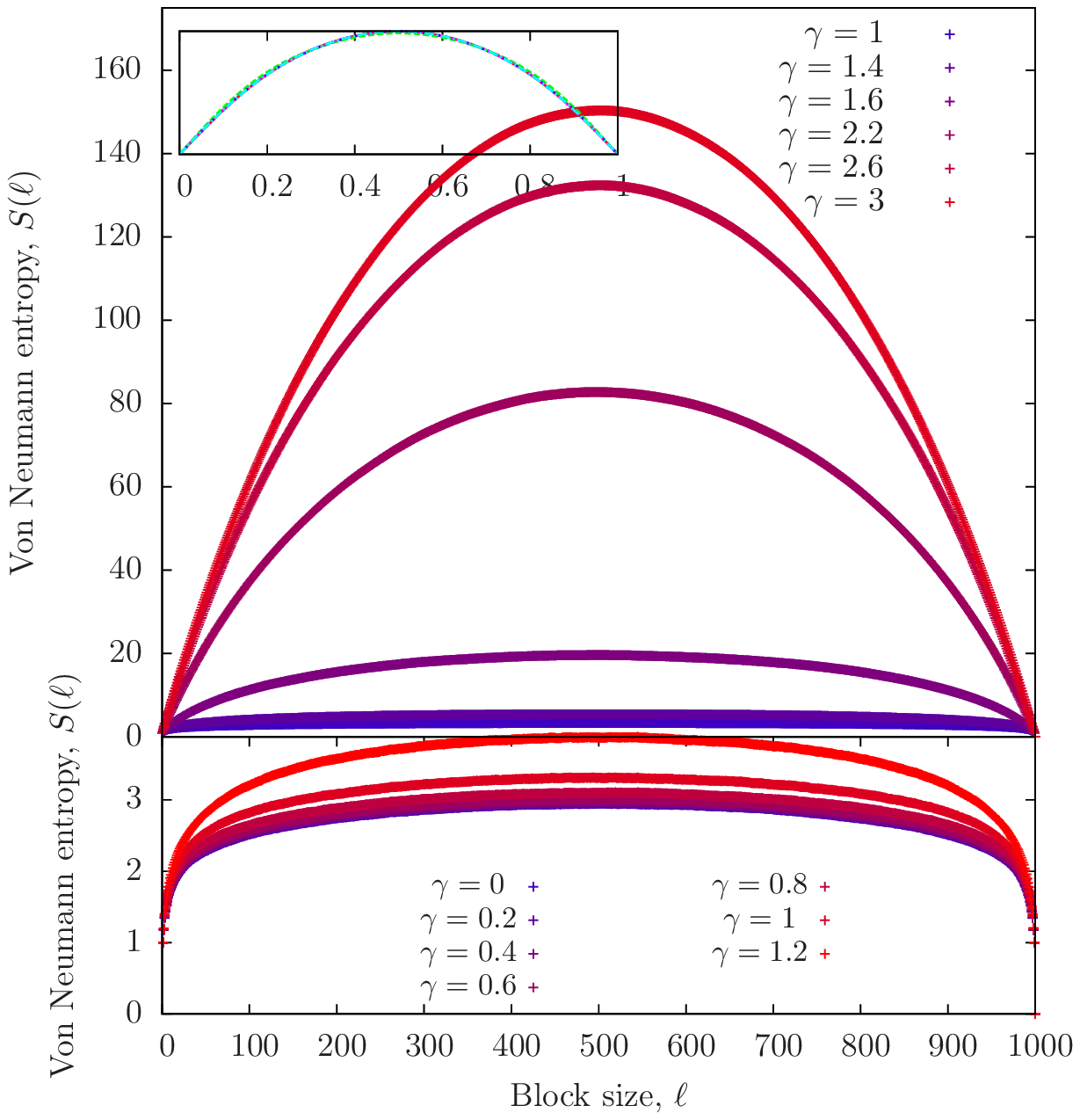,width=8cm}
\epsfig{file=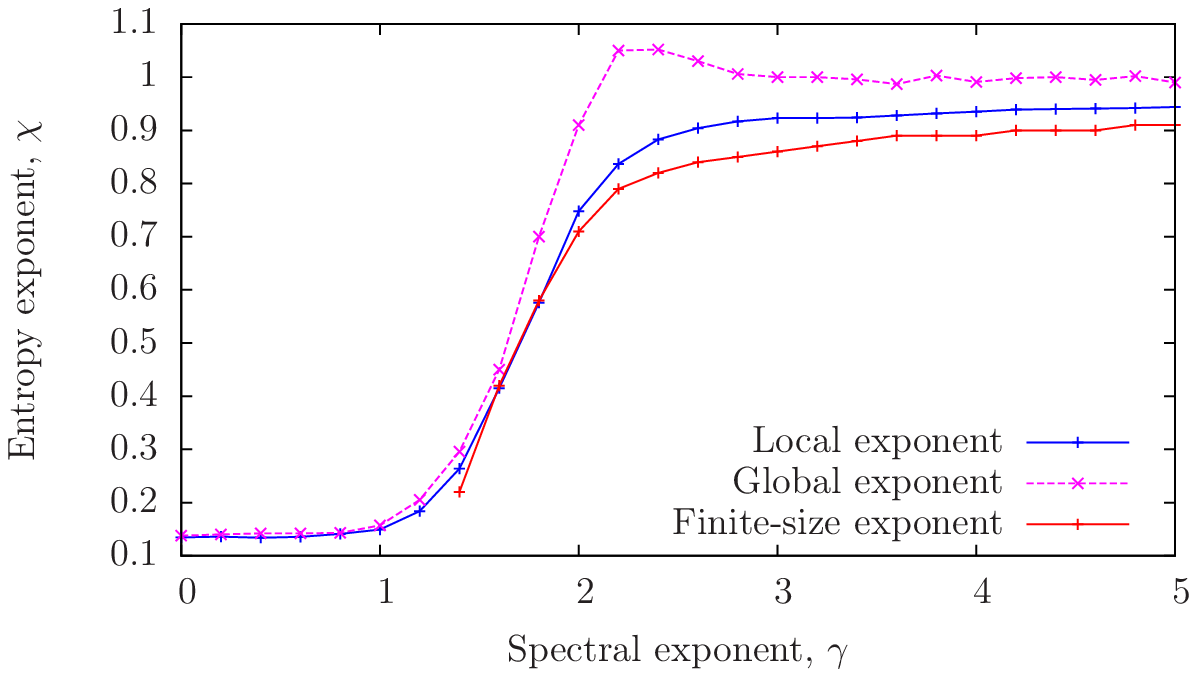,width=8cm}
\caption{(A) Average block von Neumann entropy as a function of the
  block size for different values of $\gamma$, with $N=1000$. Top
  panel: $\gamma\geq 1$. Bottom panel: $\gamma\leq 1$. Inset: scaling
  function, $Y(x)$ as in Eq. \eqref{eq:scaling_function}. (B) Fitting
  exponent $\chi$ as a function of $\gamma$ for the three strategies
  discussed in the text.}
\label{fig:entropy}
\end{figure}

The average entanglement entropy as a function of the block size
$\ell$, for a fixed value of $N=1000$, $10^5$ realizations and several
values of $\gamma$ is shown in Fig. \ref{fig:entropy} (A). The upper
part of the panel is devoted to $\gamma\geq 1$, while the lower one
shows more detail for $\gamma\leq 1$. Notice that, for $\gamma\leq 1$,
the function $S(\ell)$ is nearly independent of $\gamma$. We propose a
finite-size fit of the form:

\begin{equation}
S(\ell) \approx A \( N\, Y(\pi\ell/N) \)^\chi,
\label{eq:s_form}
\end{equation}
where the scaling function $Y(x)$ is determined via a Fourier series
expansion in the same line as \cite{Fagotti.11,Ramirez.14}:

\begin{equation}
Y(x)=\sin(x)+\sum_{n=1}^\infty \alpha_n \sin((2n+1)x).
\label{eq:scaling_function}
\end{equation}

The best fit values of $\alpha_n$ are small and nearly independent of
the spectral exponent $\gamma$. We have found $\alpha_1 \approx 0.05$
and $\alpha_2 \approx 0.005$, both slowly decaying as $\gamma$
increases. The inset of the top panel of Fig. \ref{fig:entropy} shows
these scaling functions for different values of $\gamma$.

The values of the exponent $\chi$ present more relevance. Fig
\ref{fig:entropy} (B) show these exponents, found by three different
strategies: (i) {\em Finite-size}, using a full fit to expression
\eqref{eq:s_form} for $N=1000$, (ii) {\em Local exponent}, fitting the
entropy for small blocks to a form $S(\ell) \approx \ell^\chi$ also
for $N=1000$, (iii) {\em Global exponent}, fitting $S(N/2)$ to a form
$N^\chi$ for different values of $N$, up to $N=2000$. The three
expressions differ slightly for larger $\gamma$, although they keep a
general trend: for $\gamma\leq 1$, $\chi$ is very close to zero, while
for $\gamma\to\infty$ we see $\chi\to 1$. This signals a {\em
  volumetric} growth of the entropy $S\sim \ell$. The discrepancies
between the values of $\chi$ measured by the different strategies, as
seen in Fig. \ref{fig:entropy} (B), may be of numerical origin. 

Interestingly, for $\gamma\leq 1$, the $S(\ell)$ curves are nearly
identical, and the best finite-size fit to the whole function is {\em
  not} given by the power-law expression \eqref{eq:s_form}, but
expression \eqref{eq:s_rxx}, i.e. a logarithmic behavior. Even for
$\gamma\approx 1$, the best fit is logarithmic, but with a slightly
larger prefactor. It is difficult to determine whether there is a
smooth crossover between $\gamma=0$ and $\gamma\to\infty$ or a sharp
transition at $\gamma\approx 1$, below which the entropy grows
logarithmically, i.e.: if the IRFP extends to the region $\gamma\leq
1$.

\begin{figure}
\epsfig{file=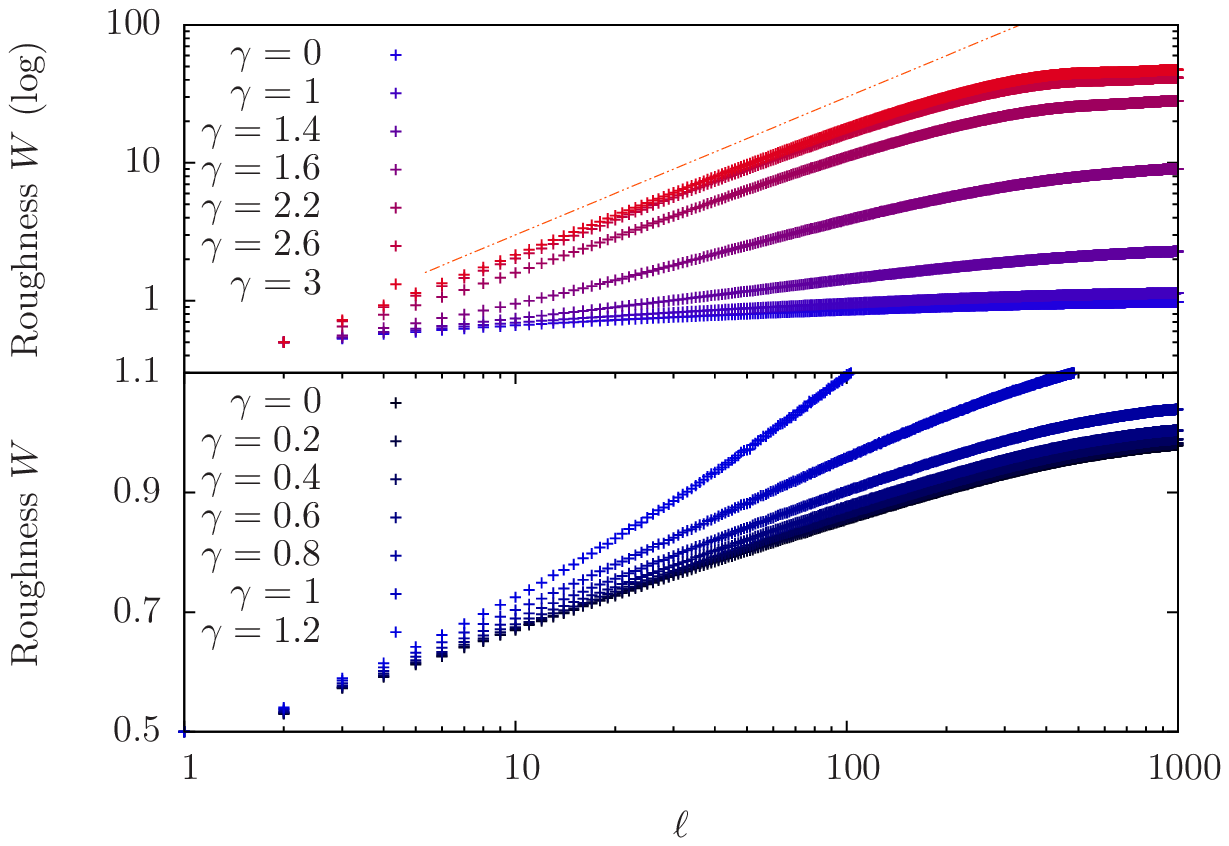,width=8cm}
\epsfig{file=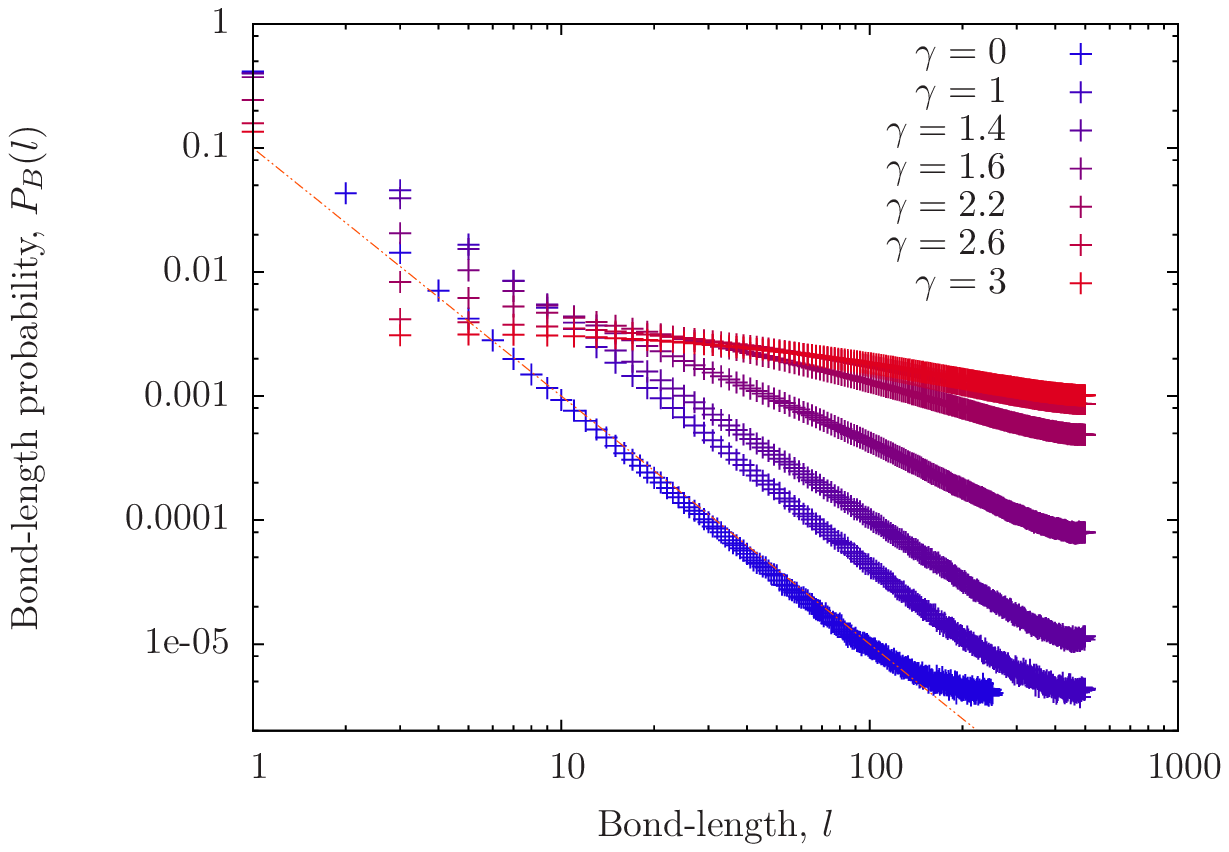,width=8cm}
\epsfig{file=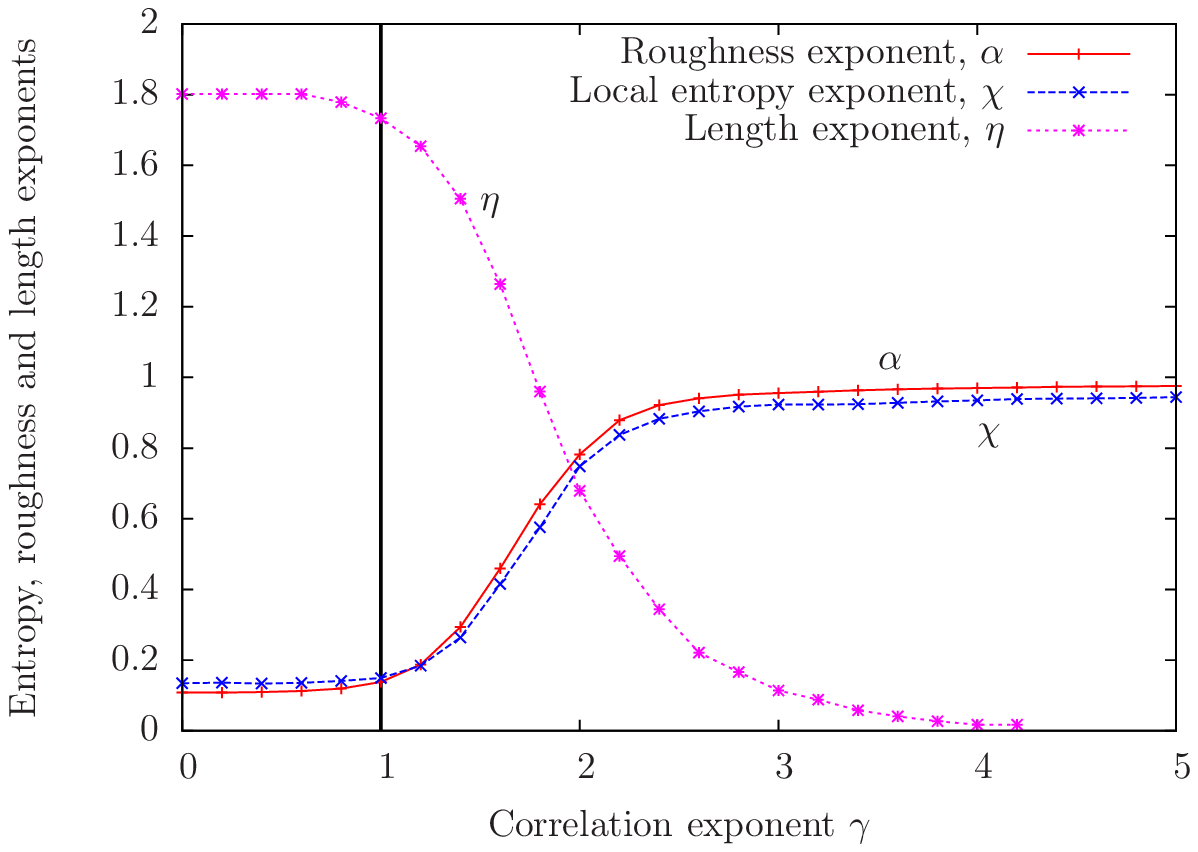,width=8cm}
\caption{(A) Roughness of blocks, $W(\ell)$, of different sizes, for
  $N=1000$ and different values of $\gamma$. The top frame is in
  log-log, while the bottom one only the $x$-axis is logarithmic. This
  way we can show that for $\gamma\leq 1$, the best fit is for a
  logarithmic growth of the roughness. (B) Bond-length distribution,
  also for $N=1000$, using the minimal length. Notice that the lines
  are parallel for $\gamma=0$ and $\gamma=1$, but shifted. They have
  the same scaling exponent, but different prefactor. (C): fitting
  exponents for the entropy ($\chi$), roughness ($\alpha$) and
  bond-length ($\eta$) as a function of $\gamma$. Below the $\gamma=1$
  line, it can be argued that the roughness and the entropy do not
  behave like a power law, but instead they show logarithmic
  behavior.}
\label{fig:roughness}
\end{figure}

Another interesting observable is provided by the study of the height
function which characterizes the state, given by
eq. \eqref{eq:height_function}. As we will show, the profiles are
fractals, of similar nature to the ones appearing in the study of
rough interfaces \cite{Family.85,Barabasi.95}. Let us define the
roughness, or width $W$, of the interface for a given length scale
$\ell$ as the average deviation of the heights in windows of that
size. Then, the Family-Vicsek Ansatz assumes that
$W\sim\ell^\alpha$. Fig. \ref{fig:roughness} (A) shows the roughness
as a function of the window size $\ell$, taking $10^5$ realizations
for each value of $\gamma$. The top frame shows a log-log plot, while
in the bottom one only the $x$-axis is logarithmic. The difference is
notorious: for $\gamma>1$, the roughness follows a clear power law,
with exponent $\alpha$ which grows up to one (shown as a straight
line). For $\gamma\leq 1$, instead, the behavior is better fit by a
logarithmic function $W(\ell)\sim \log(\ell)$. This provides further
support to the conjecture that the behavior for $\gamma\leq 1$
corresponds to the IRFP.

Panel (B) of Fig. \ref{fig:roughness} depicts the probability
distribution for the bond-length. A power-law is established, i.e.,
$P_B(l) \sim l^{-\eta}$, and $\eta$ is shown to depend on
$\gamma$. For $\gamma\leq 1$, the curves appear to be parallel, i.e.,
show the same exponent, and only differing in their
prefactor \cite{foot:renormalized}.

Fig. \ref{fig:roughness} (C) shows the values of the three exponents,
entropy ($\chi$), roughness ($\alpha$) and bond-length distribution
($\eta$) as a function of the correlation parameter $\gamma$. Notice
that $\chi$ is very similar to $\alpha$, as suggested by relation
\eqref{eq:entropy_height} which links the block entropy to the height
fluctuations. Both exponents grow with $\gamma$, starting near zero
for uncorrelated spin chains and saturating at a value close to 1. The
bond-length exponent $\eta$ behaves in the opposite way, starting at
$\eta=2$ for uncorrelated spin chains and decreasing towards zero. The
region for $\gamma\leq 1$ is peculiar: while the bond-length $\eta$
exponent is still $2$, the other two exponents are very close to zero,
since the true behavior is expected to be logarithmic.

The limit $\gamma\to\infty$ is also rather special. A look at the last
panel of \ref{fig:diagrams} shows that {\em rainbow-like} structures
become more and more prominent. The limit in which only the lowest
momentum modulation survives gives rise to a perfect rainbow state,
which presents volumetric entanglement
\cite{Vitagliano.10,Ramirez.15}, i.e., $S\sim \ell$. This explains the
limit $\chi\to 1$ for the entropy exponent for large
$\gamma$. Similarly, the height function becomes a nearly perfect
wedge, which explains the $\alpha\to 1$ behavior. In that extreme, the
bond-length distribution is completely flat, since all bond-lengths
show up once for each realization, thus $\eta=0$.


\section{RNA folding and Spin Chains}
\label{sec:rna}

As it was briefly discussed in the introduction, planar pairings also
appear naturally in the study of the secondary structure of folded RNA
strands \cite{Tinoco.99}. The model developed by Wiese and coworkers
\cite{Wiese.06,Wiese.08} works in the following way: (1) a pair of
sites with different parity are chosen randomly and paired; (2)
further pairs are chosen in the same way, always under the constraint
that no previous bonds can be crossed. In their seminal work
\cite{Wiese.08}, the authors studied the roughness of the equivalent
height function and the bond-length distribution, showing that they
both follow a power-law behavior, $W\sim \ell^\alpha$ and $P_B(l)\sim
l^{-\eta}$. Then they proved that $\alpha+\eta=2$. If we assume the
scaling equivalence of the roughness and the entropy, this result is
also fulfilled in uncorrelated random spin chains, where we have
$\alpha=0$ (because of the logarithmic behavior of the entropy) and
$\eta=2$ \cite{Refael.04,Ramirez.14}. On the other hand, this relation
does not hold for correlated spin chains.

We may ask what is the range of validity of the relation $\chi+\eta=2$
(or $\alpha+\eta=2$). Extending the results of
\cite{Hoyos.07,Ramirez.14} we can provide a proof of that statement in
the case of uncorrelated bonds. Indeed, let us consider a block of
size $\ell$ and let us number the sites from $1$ to $\ell$. The bond
at site $i$ will be cut by the block if it goes left and its length is
larger or equal than $i$, or if it goes right and its length is larger
than $\ell-i$. So, we have an estimate for the average entropy:

\begin{equation}
S(\ell) \approx \sum_{i=1}^\ell {1\over 2}\( P_B(l\geq i) +
P_B(l>\ell-i) \) = \sum_{i=1}^\ell P_B(l\geq i)
\label{eq:s_prob}
\end{equation}

This equation implies a double integration. If $P_B(l)\sim l^{-\eta}$,
it leads to $S(\ell) \sim l^{-\eta+2}$, as we desired. As it follows
from Fig. \ref{fig:roughness} (C), this is not true for the planar
state ensembles generated with correlated couplings. In fact, in the
{\em rainbow} limit, we have $\chi+\eta\to 1$, which suggests a strong
correlation between the bonds.

\subsection{The inverse problem}

How strong is the connection between the RNA folding and disordered
spin chains? Can we obtain an ensemble of couplings $\{J_i\}$ such
that the ground states of Hamiltonian \eqref{eq:ham} correspond to the
planar states obtained in RNA folding? This question leads us to the
study of the more general {\em inverse SDRG problem}.

If we regard the SDRG as a mapping between sets of couplings and
planar pairings, we might be able to reverse the algorithm, and obtain
the set of couplings which give rise to a certain planar pairing. In
other terms, a parent 1D Hamiltonian for a given planar state. In this
section we will show that (1) every planar state has a (non-unique)
parent 1D Hamiltonian and (2) an explicit algorithm to obtain the {\em
  optimal} set of couplings, in a sense to be determined later.

\begin{figure*}
\epsfig{file=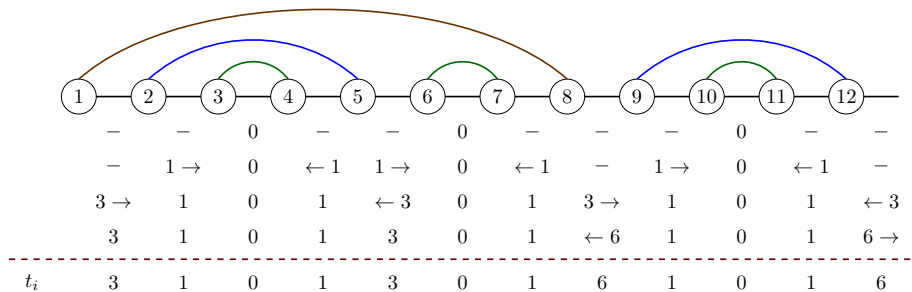,width=12cm}
\caption{Illustration of the inverse algorithm to obtain the optimal
  log-couplings which give rise, via the SDRG, to a given planar bond
  structure. Each row corresponds to one of the steps of the
  algorithm. The log-couplings which have not been assigned yet appear
  as a ``$-$''. The arrows which appear close to a new value point to
  the bond which has created (or modified) that value.}
\label{fig:illust_inverse}
\end{figure*}

The aim is to obtain the logarithmic couplings, $\{t_i\}$, given the
set of bonds, $\{{\bf p}_i\}$. Our proposed algorithm works as follows
(see Fig. \ref{fig:illust_inverse} for an illustration):

\begin{itemize}
\item{} Sort the bonds in order of increasing length.
\item{} Consider the bonds of length one, fix their internal
  log-couplings to $0$. In the first row of
  Fig. \ref{fig:illust_inverse}, we put a zero under links $3-4$ and
  $10-11$.
\item{} Flank these zeroes with log-couplings of value $1$ at both
  sides. See the second row of Fig. \ref{fig:illust_inverse}, where
  the arrows in the new values point to the zero which they flank.
\item{} Now consider the bonds of length three. Find the effective
  log-coupling which would appear as their {\em renormalization value}
  (which must be 2). Flank them with log-couplings of value $2+1=3$ at
  both sides, as in the third row of Fig. \ref{fig:illust_inverse}.
\item{} Consider the rest of the bonds in order of increasing
  lengths. For each of them, find their {\em renormalization value}
  and flank them with log-couplings of value one unit higher.
\item{} Log-couplings may never decrease along the procedure. If two
  values collide, take the larger.
\end{itemize}

This procedure yields couplings which, by construction, always give
rise to the desired bond structure. Moreover, because the value of
each bond is computed using the SDRG itself, we ensure a certain {\em
  optimality condition}: among the sets of couplings yielding the
desired state, our choice will always require the minimal span of
coupling values. For example, this Hamiltonian will yield the largest
possible gap.

\begin{figure}
\epsfig{file=rnainverse.eps,width=8cm}
\epsfig{file=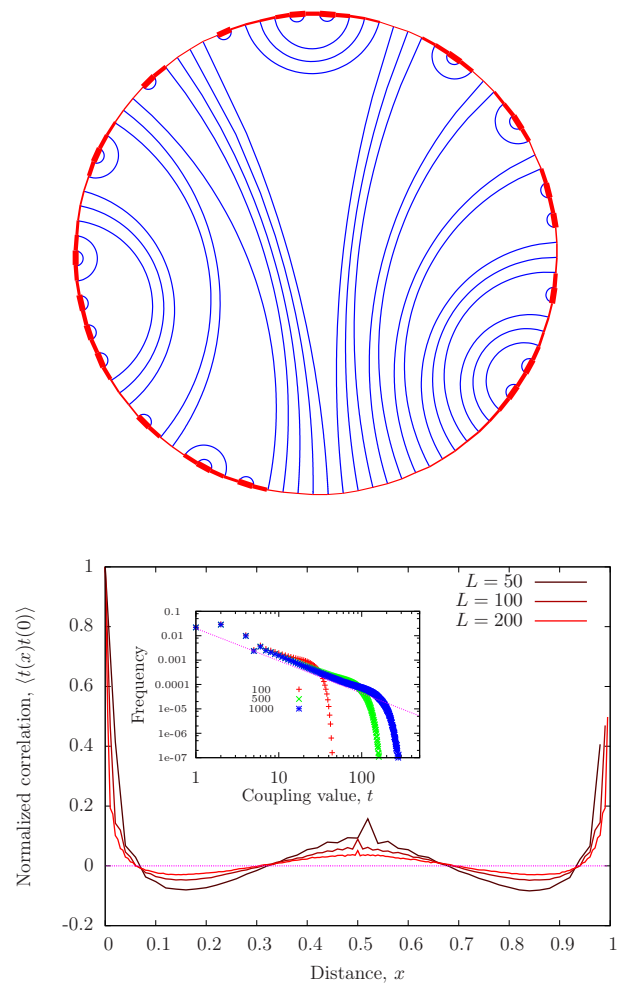,width=8cm}
\caption{(A) The optimal couplings which give rise to a given planar
  pairing obtained from the RNA folding algorithm with $N=100$. (B)
  The correlation function for the log-couplings at different points
  for $N=50$, $100$ and $200$, using $10^6$ realizations. Inset:
  histogram for the log-coupling values for $N=100$, $500$ and $1000$,
  also with $10^6$ realizations. Notice the power-law inertial range,
  with an exponent close to $-4/3$ (straight line).}
\label{fig:rnainverse}
\end{figure}

Fig. \ref{fig:rnainverse} (A) shows the couplings which give rise to a
give instance of the RNA folding problem with $N=100$. We have run
$10^5$ simulations of the RNA folding algorithm and obtained the
optimal couplings for different system sizes up to
$N=1000$. Fig. \ref{fig:rnainverse} (B) shows the (translation
invariant) correlation function for the log-couplings in the $N=50$,
$100$ and $200$ cases. The values present long range correlations, but
not a clear power-law behavior. Moreover, the couplings field
$\{t_i\}$ is {\em not} gaussian. In the inset of
Fig. \ref{fig:rnainverse} (B) we show the histogram, in logarithmic
scale, for $t_i$. The marginal probability distribution is not
gaussian. Instead, it is a power-law, with an empirical exponent close
to $-4/3$.

\section{Generic planar states}
\label{sec:generic}

Since we have determined that all planar states have a 1D parent
Hamiltonian, we may still ask how dense are planar states within the
Hilbert space. In other terms, how generic they are. We can define an
ensemble of planar states for $N$ sites under the condition that all
possible planar pairings have the same probability. In order to sample
that ensemble, we just apply a correction to the RNA folding sampling
strategy. In the RNA folding algorithm, the pair of sites $(i,j)$
which will constitute the next bond is chosen with equal probabilities
among those which do not cut any previous bond. But, following that
procedure, not all planar pairings are sampled with the same
probability. This can be corrected if the probabilities for each pair
$(i,j)$ are not equal, but proportional to the number of planar
pairings which are consistent with the presence of that bond.

Let us consider a certain {\em empty patch} of length $n$ in a planar
pairing which is under construction, i.e., a set of contiguous spins
which have not been paired yet. As we know, there are $P_n$ possible
ways to create a planar pairing on that empty patch. The spin with
index 1 must be paired with some spin inside the patch, let us refer
to its index as $k$. Then, after bond $(1,k)$ is established, the
number of different possible planar pairings will be
$P_{k-2}P_{n-k}$. Thus the probability with which bond $(1,k)$ should
be taken is just $P_{k-2}P_{n-k}/P_n$, which is known to be less than
one by construction, as we see in Eq. \eqref{eq:recurrence}. Repeating
this procedure, we can sample the planar pairing ensemble with equal
probabilities.

We have found numerically the average block entropy as a function of
the block size $\ell$ for this ensemble of states, and found that it
grows as $S(\ell)\sim \ell^\chi$, with $\chi\approx 0.54$. The precise
value is not very relevant, but it allows us to conclude that planar
states are highly non-generic quantum states, because for generic
states we should obtain $S(\ell) \sim \ell$, i.e., a volumetric growth
of the entropy.


\section{Conclusions and Further Work}
\label{sec:conclusions}

In this article we have applied the SDRG to study the ground state
properties of a strongly disordered random spin chain with long-range
correlations between its couplings. The states can be described as
valence bond states with planar bond structures, and they can have
arbitrarily large entanglement entropy. Concretely, we have chosen the
couplings such that their logarithm is expressed as a Fourier series
with random coefficients, falling as a power-law of the momentum
$k^{-\gamma}$. For $\gamma\leq 1$ the behavior is very similar to the
infinite randomness fixed point (IRFP) found for uncorrelated coupling
constants. Nonetheless, for $\gamma>1$, the block entropy behaves as a
power-law of the block size, $S\sim\ell^\chi$, with $\chi$ a function
of the exponent $\gamma$ which seems to interpolate smoothly between
$\chi=0$ and $\chi=1$ as $\gamma\to\infty$. The bond length
probability, which is related to the correlator, is also characterized
by a power-law, $P_B(l)\sim l^{-\eta}$, with $\eta=2$ for $\gamma\leq
1$ and falling to $\eta\sim 0$ for $\gamma\to\infty$. This extreme,
$\gamma\to\infty$, corresponds to the case where only the lowest
momentum $k=2\pi/N$ contributes to the correlation between the
couplings, and the state becomes a {\em rainbow state}. As we have
shown, the planar states can be mapped to a 1D interface, whose
roughness behaves approximately like the entanglement entropy, as it
is suggested by expression \eqref{eq:entropy_height}. Remarkably, the
system described constitutes a family of local 1D Hamiltonians whose
ground states violate the area law to any desired degree.

We have also considered the inverse renormalization problem: given a
(planar) valence bond state, to obtain its (1D) parent Hamiltonian. In
this way we were able to study the ensemble of random spin chains
whose ground states would correspond to the planar structures which
show up in other physical situations, such as the RNA folding
problem. These {\em engineered} random spin chains present a behavior
of the entanglement entropy and the correlators which do not
correspond to any value of $\gamma$. This suggests that the phase
diagram of random spin chains with large correlations between the
couplings is richer than expected.

Inhomogeneous spin chains can be mapped, in some cases, to models
which represent the motion of fermionic matter on a curved spacetime
\cite{Boada.11}, where the metric is given by the coupling
constants. Thus, our study shows that the statistical properties of
the metric show up as statistical properties of the entanglement of
the vacuum, i.e., the ground state of the corresponding
Hamiltonian. Moreover, we can also find, using the inverse
renormalization algorithm, the optimal spatial geometry which gives
rise to a certain vacuum entanglement. These results may shed light on
the relation between entanglement and space-time \cite{Maldacena.13}.

\begin{acknowledgments}
We would like to thank J. Cuesta for insights into the statistical
mechanics of RNA folding. This work was funded by grants
FIS-2012-33642 and FIS-2012-38866-C05-1, from the Spanish government,
QUITEMAD+ S2013/ICE-2801 from the Madrid regional government and
SEV-2012-0249 of the ``Centro de Excelencia Severo Ochoa'' Programme.
\end{acknowledgments}


\end{document}